# Intrinsic Defect Formation in Peptide Self-Assembly

Li Deng(邓礼),[1] Yurong Zhao(赵玉荣),[1] Hai Xu(徐海),[1,a] and Yanting Wang(王延颋)[2,b]

[1]*Centre for Bioengineering and Biotechnology, China University of Petroleum (East China), 66 Changjiang West Road, Qingdao 266580, People's Republic of China*

[2]*State Key Laboratory of Theoretical Physics, Institute of Theoretical Physics, Chinese Academy of Sciences, 55 East Zhongguancun Road, P. O. Box 2735 Beijing, 100190 China*

In contrast to extensively studied defects in traditional materials, we report here for the first time a systematic investigation of the formation mechanism of intrinsic defects in self-assembled peptide nanostructures. The Monte Carlo simulations with our simplified dynamic hierarchical model revealed that the symmetry breaking of layer bending mode at the two ends during morphological transformation is responsible for intrinsic defect formation, whose microscopic origin is the mismatch between layer stacking along the side-chain direction and layer growth along the hydrogen bond direction. Moreover, defect formation does not affect the chirality of the self-assembled structure, which is determined by the initial steps of the peptide self-assembly process.

[a] Author to whom correspondence should be addressed. Electronic mail: xuh@upc.edu.cn
[b] Author to whom correspondence should be addressed. Electronic mail: wangyt@itp.ac.cn

Because of thermal fluctuations, defects usually form during the growth process of traditional materials, such as grapheme,[1] crystalline structures of ZnO,[2] and liquid crystals.[3] Although defects should be avoided to produce materials with a high purity, in some other cases they are desired because they can endow the materials with unique and useful electronic, optical, thermal, and elastic properties. For instance, the efficient applications of semiconductors are based on our knowledge of their defects.[4]

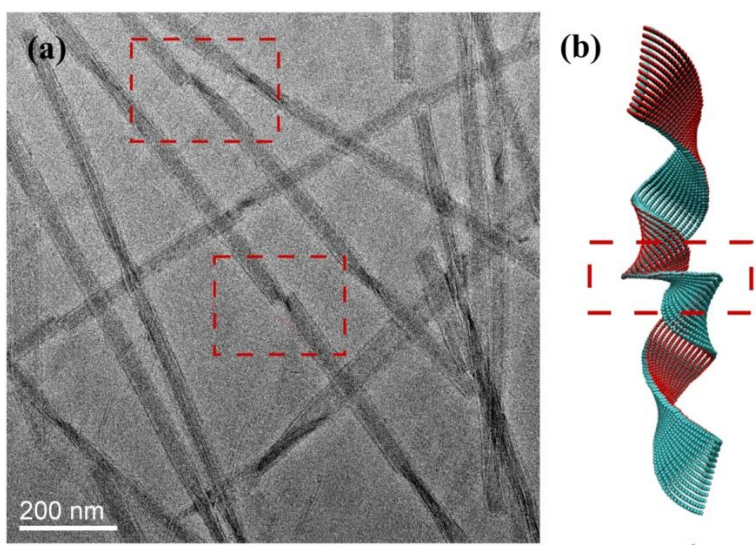

FIG. 1. Defects in self-assembled KI4K nanotubes observed in the TEM experiment (a) as well as formed in the MC simulation (b).

Beyond the scope of traditional materials, biomaterials formed by self-assembly provides novel functionalized materials as well as new knowledge to chemistry, biology, and medicine,[5,6] among which peptide self-assembly not only is used as the template for developing novel low-dimensional nanomaterials,[7,8] but also plays a central role in the formation of some neurodegenerative diseases.[9,10] Studied by various experimental techniques including fiber-XRD, NMR, TEM, and AFM, peptides are known to be able to self-assemble into ordered structures, such as nanofibrils, nanoribbons, and nanotubes, through forming the primary structure of cross β tapes[11-13] and layer stacking of tapes onto the primary structure along the peptide side-chain direction.[14,15]

In the light of their analogue in traditional materials, it is natural to infer that defects also play an essential role in biomaterials. Surprisingly, in contrast to the extensive studies on self-assembled morphologies of peptides, no studies so far have been devoted to their defects except a few marginal experimental observations.[16,17] Ulijn et al.[16] observed ruptured ribbon structures in the hydrogel self-assembled by Fmoc-FF under physiological conditions and attributed it to the torture by shear forces. In our previous work,[17] we observed a few defected nanotubes self-assembled by KIIIIK (KI4K), which inspired our interest in systematically investigating the microscopic mechanism of defect formation in peptide self-assembly.

In this study, with special attention to defects, we performed again the KI4K self-assembly experiment, but now adding in the aqueous solution some methanol, which is known by our experience to increase the probability of defect formation. The experimental setup was the same as described in Ref. 17. When 40% methanol was added in the aqueous solution, some defected nanotubes formed, whose TEM picture is shown in Fig. 1(a). It can be seen that the defects do not look like formed through the torture by shear forces, rather more likely to form intrinsically during the dynamic process of self-assembly, since the nanotube structures above and below a defect are regular and the internal structure of the defect itself is still well organized. In order to understand its microscopic mechanism, which does not necessarily resemble the defect formation mechanism during the growth of traditional materials, we developed a simplified model to simulate the self-assembly process forming intrinsic defects, as described below.

Various simplified models[18-25] have been constructed to simulate peptide self-assembly, whose large temporal and spatial scales are still far beyond the reach of all-atom molecular simulations. The coarse-grained models[18-20] have unravelled the microscopic mechanism of cross-β tape formation, but still contain too many degrees of freedom to study the mesoscopic morphology. The static elasticity models[21,22] cannot study the microscopic mechanism of the morphology transformation dynamics in peptide self-assembly but reveal the relation between morphology and the width, which have been confirmed by recent experiments.[26] Selinger et al.[23] developed an elastic model to study morphological transformations of chiral molecules, which is however unsuitable for this study due to the lack of direct mapping between their molecular model and the peptide molecular structure. Aggeli's model[24,25] considered the peptide molecular structure appropriately, but it was designed to study the static conformations and cannot be used to study morphology transformation dynamic processes. Based on both Selinger's and Aggeli's models, we managed to develop a dynamic hierarchical simplified model specifically designed for studying the intrinsic defect formation mechanism during morphology transformation in peptide self-assembly processes.

In our hierarchical model, a peptide molecule is represented by a rod with three characteristic directions along the peptide backbone, the hydrogen bond (H-bond) growth, and the side-chain layer stacking, respectively. As shown in Fig. 2(a), the three directions are perpendicular to each other, so they are used to construct the local coordinate $(\vec{S}, \vec{P}, \vec{H})$. Peptide molecules are bounded together by H-bond interactions to form a twisted layer, in which neighboring peptides have both stretching and twisting interactions, as illustrated in Fig. 2(b). As shown in Fig. 2(c), layer stacking due to the hydrophobic interaction between layers increases the width of the self-assembled structure.

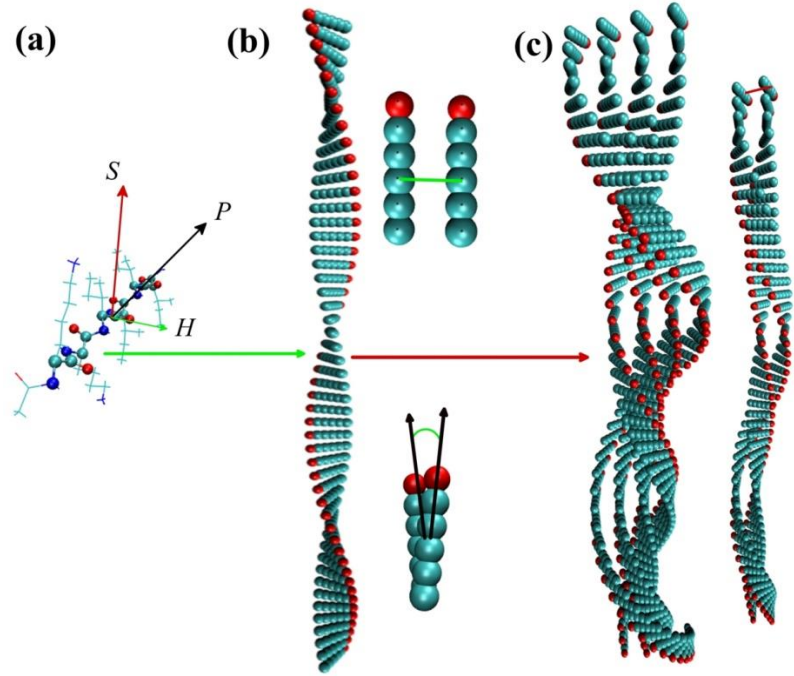

FIG. 2. Schematic of the dynamic hierarchical model for peptide self-assembly. (a) Local coordinates on a peptide molecule. The black arrow represents the backbone direction, red the layer-stacking direction, and green the H-bond growth direction. (b) Formed layer structure (left) and illustration of the stretching energy (upper right) and the twisting energy (lower right) between two neighboring peptides in the same layer. (c) Stacked layers with one end of peptide molecules colored with red to guide the eyes.

Correspondingly, the energy terms of our model are:

$$E = \sum_{I=1}^{N} \sum_{i,j=1}^{n} [k_1(D_{ij}^I - D_0)^2 + k_2(R_{ij}^I - R_0)^2 + k_3(\theta_{ij}^I - \theta_0)^2] - (N-1) \times \sigma_s \quad (1)$$

where $N$ is the number of layers and $n$ is the number of peptides in one layer. The three terms in the square brackets describe the intralayer interactions between neighboring peptides, namely stretching along the H-bond direction (first term), stretching perpendicular to the H-bond direction (second term), and twisting elastic energy (third term), all described by harmonic potentials with $k_1$, $k_2$, and $k_3$ the elastic constants and $D_0$, $R_0$, and $\theta_0$ the equilibrium constants. Two different stretching terms in our model well represent the anisotropy of the stretching motion. The third term describes the twisting energy cost between neighboring peptides away from their equilibrium angle. The last term describes the total interlayer energy with the approximation that the attraction between two neighboring layers is a constant $\sigma_s$. Similar to Aggeli's model,[24,25] our model assumes that the self-assembly process starts from a primary layer with a cross β structure, and the peptide positions in a new layer are determined by the layer they attach to.

In compliance with the experimentally discovered pathway for self-assembly,[27,28] our Monte Carlo (MC) simulation associated with the above model was designed to include two kinds of trial moves at different time scales, whose steps are: (1) an equilibrated layer structure with $n$ peptides is constructed as the primary layer; (2) a trial move for the existing peptides at every step is attempted according to the Metropolis algorithm;[29] (3) a trial stacking of a new layer at every $5\times10^4$ steps is attempted according to the Metropolis algorithm to simulate the width growth in peptide self-assembly; (4) the above trial moves are repeated for $5\times10^8$ steps. Referring to regime 4 of Aggeli's model[24,25] when peptides self-assemble into a twisted ribbon with finite layers, we set the system parameters to $k_1=k_2=2$, $k_3=4$, $\sigma_s=25$, $\theta_0=0.12$ rad, $D_0=R_0=0.5$, $h_0=1.6$ and $n=200$ and the free boundary condition was applied to both ends of the structure. The system temperature was as low as 0.01 to allow the simulation essentially an energy minimization procedure. However, in our MC simulations, the primary layer is allowed to deform into other structures to enable morphology transformation when the width changes, which is the major difference from Aggeli's model.

Consistent with the experimental observation that the width growth becomes slower as time evolves,[28] the attachment of new layers becomes slower in our simulation, as shown in Fig. 3(a). Fig. 3(b) demonstrates that the initial energy (energy barrier for attachment) is larger for a later-joined layer, so the final structure can only contain a finite number of layers. Moreover, before the system is equilibrated, the outer layers always have larger energies than the inner layers, suggesting that the outside part of the system plays an important role in the formation of different morphologies.[23]

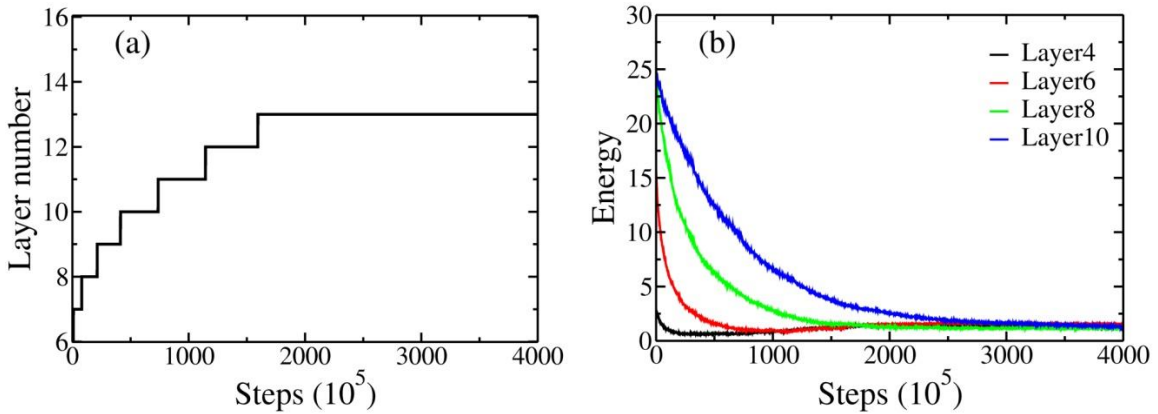

FIG. 3. Layer number (a) and energies of different layers (b) versus simulation time.

Fig. 1(b) shows a typical simulated structure with a defect formed in the middle, which resembles the experimental structure shown in Fig. 1(a). The central axes of the two parts above and below the defect do not align with each other, but their chirality is the same. The simulated structure was quantified by the Gaussian curvature defined as[23,30]

$$\kappa = 2(2\pi - \sum_{i=1,4}\delta_i)/\sum_{i=1,4}a_i \qquad (2)$$

where the sums are over four adjacent sites (the center-of-mass positions of peptides in our case), $\delta_i$ is the angle between neighboring bonds connecting the sites, $a_i$ is the area of the triangle adjacent to the site. The calculated Gaussian curvature averaged over all layers for the structure in Fig. 1(b) is plotted in Fig. 4(a). The defect in the middle has a large curvature, while the rest part of the tube has a curvature around zero.

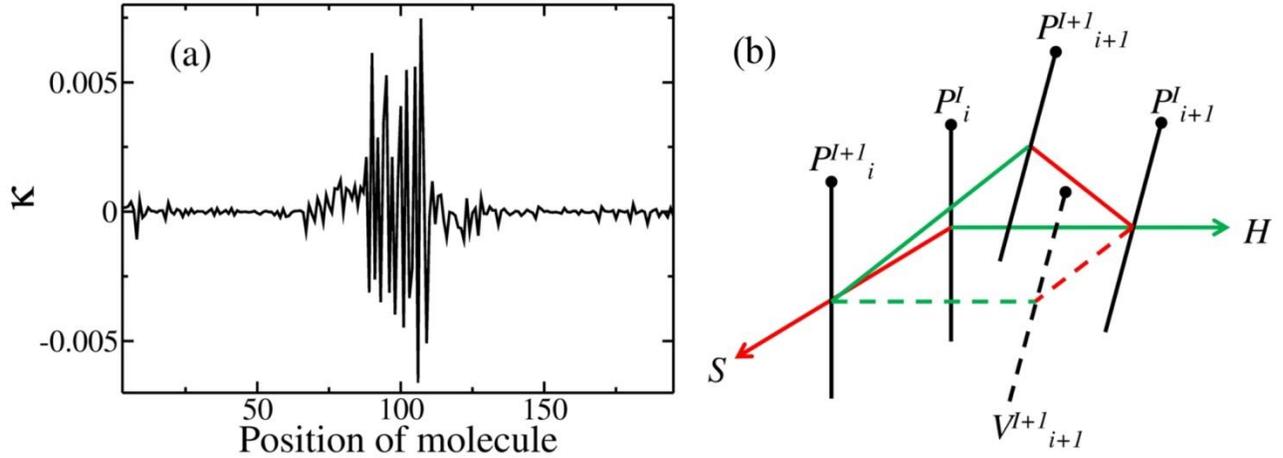

FIG. 4. (a) Gaussian curvature averaged over all layers for the simulated structure shown in Fig. 1(b). (b) Schematic illustration of the mismatch between layer stacking along the side-chain direction and peptide growth along the H-bond direction.

The energy of each layer in Fig. 3(b) demonstrate that large deformation energies of outer layers result in the morphological transformation from a twisted fibril to a tube, whose microscopic origin is the mismatch between layer stacking along the side-chain direction and peptide growth along the H-bond direction, schematically illustrated in Fig. 4(b). Because two neighboring peptides in layer $I$, denoted as $P_i^I$ and $P_{i+1}^I$, are aligned with a twisting angle, the newly-attached peptide $P_{i+1}^{I+1}$ in layer $I+1$ adjacent to $P_{i+1}^I$ should be in the position

$$\vec{R}_{i+1}^{I+1} = h_0 \cos\theta_0 \vec{S} + h_0 \sin\theta_0 \vec{P} + D_0 \vec{H} \qquad (3)$$

to reduce the hydrophobic surface. On the other hand, energy minimization of layer $I+1$ requires the same peptide to be in a virtual position denoted by $V_{i+1}^{I+1}$ in Fig. 4(b), which induces a force $F_{i+1,\,i}^{I+1}$ along the vector

$$\Delta\vec{R}_{i+1,i}^{I+1} = h_0(\cos\theta_0 - 1)\vec{S} + h_0 \sin\theta_0 \vec{P} + D_0 \vec{H} \qquad (4)$$

Another force $F_{i-1,\,i}^{I+1}$ between $P_{i-1}^{I+1}$ and $P_i^{I+1}$ is opposite to $F_{i+1,\,i}^{I+1}$ when $\theta_0$ is small. Initially, since the peptides in between have two opposite forces cancelled, only the peptides at both ends bear large forces. When they adjust their positions to minimize the forces, the adjacent peptides lose their force balance and also have to adjust their positions. Through this process, the large forces applied to the two ends propagate into the middle. When enough number of layers accumulates adequate stress, the whole morphology transforms from a fibril to a tube (Fig. 5).

Defects form intrinsically during the above morphological transformation process. Fig. 4(b) shows that the force direction at the left end is $(-\vec{S}, \vec{P}, \vec{H})$ and the right is $(-\vec{S}, -\vec{P}, -\vec{H})$, so the layer bending modes at the two ends are opposite. As shown in Fig. 1(b), the bending direction below the defect is along $\vec{P}$ with the red inner face, while above the defect it is along $-\vec{P}$, so the red side becomes the outer face. The bending directions at both ends do not agree with each other, leading to the intrinsic defect when the two bending modes meet in the middle.

Note that in our MC simulations, the layers are joined from one side and only the primary layer has its shape adjusted by trial moves, so the layer bending mode always has its symmetry broken and defects form deterministically. Nevertheless, in experiments, since most of the time more growth freedoms preserve the layer bending symmetry, only when thermal fluctuations break the symmetry can defects form through the simulated mechanism. Therefore, in peptide self-assembly experiments, defects form stochastically with a certain probability.

From Fig 1(b), we can know that the chirality of tube in two sides of defect is the same, and then the chirality formation and its relation with the defect formation mechanism have been studied by simulations. As shown in Fig. 5, our MC simulations demonstrate that the self-assembled morphology always has a unified chirality formed through the mechanism proposed by Weatherford and Saleme[31] that the twisting of the H-bond sites originates the initial chirality of the β-sheet structure. Nevertheless, both simulated morphologies with different chiralities form defects in the middle, indicating that the chirality formation mechanism is independent of the defect formation mechanism.

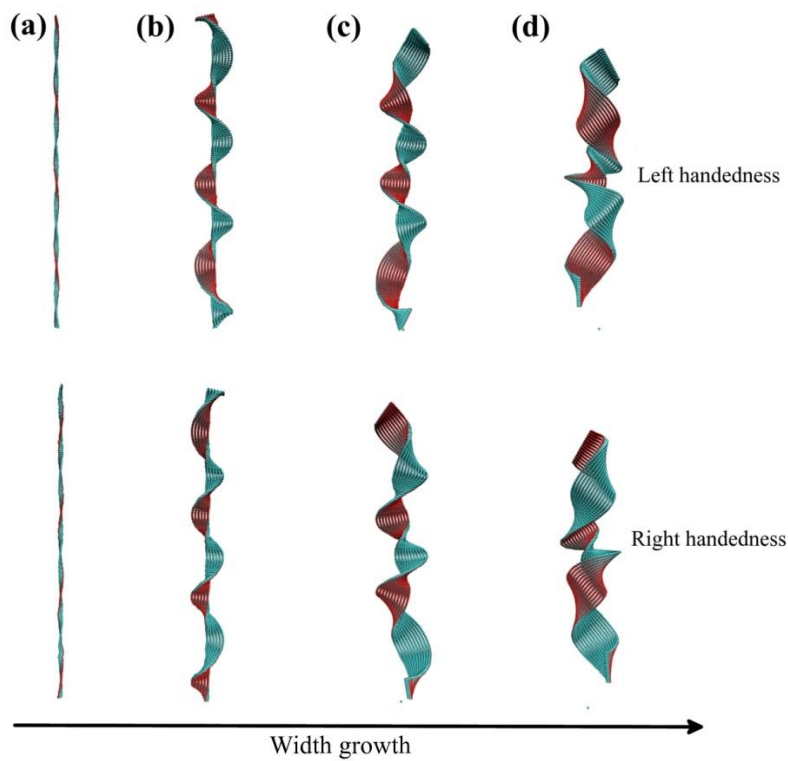

FIG. 5. Simulated peptide growth with the left chirality (upper row) and the right chirality (lower row).

In summary, we have studied in detail intrinsic defects appeared in peptide self-assembly experiments by MC simulation with a dynamic hierarchical simplified model. Our simulation results reveal that an intrinsic defect forms when the layer bending symmetry at both ends of the self-assembled structure breaks during the morphological transformation from a twisted fibril to a tube driven by the large elastic deformation of outer layers, whose microscopic origin is the mismatch between layer stacking along the side-chain direction and peptide growth along the H-bond direction. The chirality of the self-assembled structure, determined by the initial steps, is not directly related to the intrinsic defect formation. The suggested mechanism provides a theoretical guidance for future defect-related peptide self-assembly studies. For instance, it will help us to understand why adding some methanol in the aqueous solution increases the probability of intrinsic defect formation in KI4K self-assembly.

To our knowledge, this is the first systematic study of defects in biomaterials. Since many other kinds of biomaterials share common microscopic self-assembly mechanisms with peptides, our findings are expected to be helpful not only for quality control of biomaterial growth via self-assembly, but also for producing functionalized biomaterials with new features. For example, the electric resistance of a self-assembled DNA structure might be quantitatively tuned by regulating the amount of intrinsic defects formed during the self-assembly process.

## ACKNOWLEDGMENTS

This work was funded by the National Basic Research Program of China (973 program, No. 2013CB932804), the National Natural Science Foundation of China (Nos. 21373270, 91227115, and 11421063), and Application Research Foundation for Post-doctoral Scientists of Qingdao (T1404096). HX acknowledges the support by the Program for New Century Excellent Talents in University (NCET-11-0735). Allocations of computer time from SCCAS and ITP-CAS are gratefully acknowledged.